\documentclass[10pt,journal,compsoc]{IEEEtran} 
\ifCLASSINFOpdf
\else
\fi

\usepackage[space,nocompress]{cite}
\usepackage{graphicx,caption,amssymb,amsmath,array}
\usepackage{color,float}
\usepackage{subfigure,epsfig}
\usepackage{ragged2e}
\usepackage{enumitem}
\usepackage{footnote}

\usepackage{algorithm}
\usepackage{algorithmic}
\usepackage{multirow}
\usepackage{float}
\usepackage{url}

\usepackage{enumitem}
\usepackage{array}
\usepackage{booktabs}
\usepackage{bm}
\usepackage{color}

\setlist[itemize]{leftmargin=*}
\setlist[enumerate]{leftmargin=*}
\begin{document}
\title{A Benchmark Dataset for Micro-video Thumbnail Selection}

\author{Bo~Liu}

\markboth{Journal of \LaTeX\ Class Files,~Vol.~14, No.~8, August~2015}%
{Shell \MakeLowercase{\textit{et al.}}: Bare Demo of IEEEtran.cls for IEEE Transactions on Magnetics Journals}

\IEEEtitleabstractindextext{%
\justifying  
\begin{abstract}
The thumbnail, as the first sight of a micro-video, plays pivotal roles in attracting users to click and watch. Although several pioneer efforts have been dedicated to jointly consider the quality and representativeness for selecting the thumbnail, they are limited in exploring the influence of users’ interests. While in the real scenario, the more the thumbnails satisfy the users, the more likely the micro-videos will be clicked.
In this paper, we aim to select the thumbnail of a given micro-video that meets most users’ interests. Towards this end, we construct a large-scale dataset for the micro-video thumbnails. Ultimately, we conduct several baselines on the dataset and demonstrate the effectiveness of our dataset.
\end{abstract}

}

\maketitle

\IEEEdisplaynontitleabstractindextext

\IEEEpeerreviewmaketitle

\section{Introduction}
Micro-video~\cite{MMGCN,NMCL,Jiang}, as a new media type, allows users to record their daily life within a few seconds and share over social media platforms (e.g., Instagram\footnote{https://www.instagram.com/}, and Tiktok\footnote{https://www.tiktok.com/}). Considering Instagram as an example, as of June 2018, over 30 million micro-videos had been shared daily, but about 70 percent of them had not been watched. To retain users’ stickiness, beyond improving the quality of micro-videos, micro-video platforms and publishers have to draw users’ eyes quickly. 

As the most representative snapshot, the thumbnail summaries the posts and provides the first impression to the observation~\cite{1}. In order to select the thumbnail for the posts, some prior studies are presented to represent them, which can be roughly divided into two groups: monomodal-based and multimodal-based methods. In particular, the former ones~\cite{3,4,5} merely use the visual information to represent the posts. Whereas, the multimodal-based methods incorporate the visual content with the side-information, like title, description, and transcript to enhance the posts' representation learning. 

Despite the remarkable performance, we argue that they forgo the observations' comments for the posts, which reveals the observers' intents towards the posts. Specially, as shown in Figure 1, we find that some words, like “\textit{ice cream}”, “\textit{Macarons}” and “\textit{cookies}”, reflect the observers' interests to the dessert. Based on the fact that people tends to click their interesting posts, we propose to discover the observations' intents distribution from their comments and select the frame meeting the majority of observers' intents as the thumbnail of micro-video. 

To this end, we construct a large-scale benchmark dataset from Instagram platform through collecting the micro-videos associated with their thumbnails. Afterward, we capture the words corresponding to observers' intents from the comments on the platform. Finally, we obtain $18,279$ publishers, $35,148$ micro-videos, and $111$ intent words. 
To evaluate the benchmark dataset, we choose several state-of-the-art models and conduct the experiments.
\begin{figure}
	\centering
	  \includegraphics[width=0.47\textwidth]{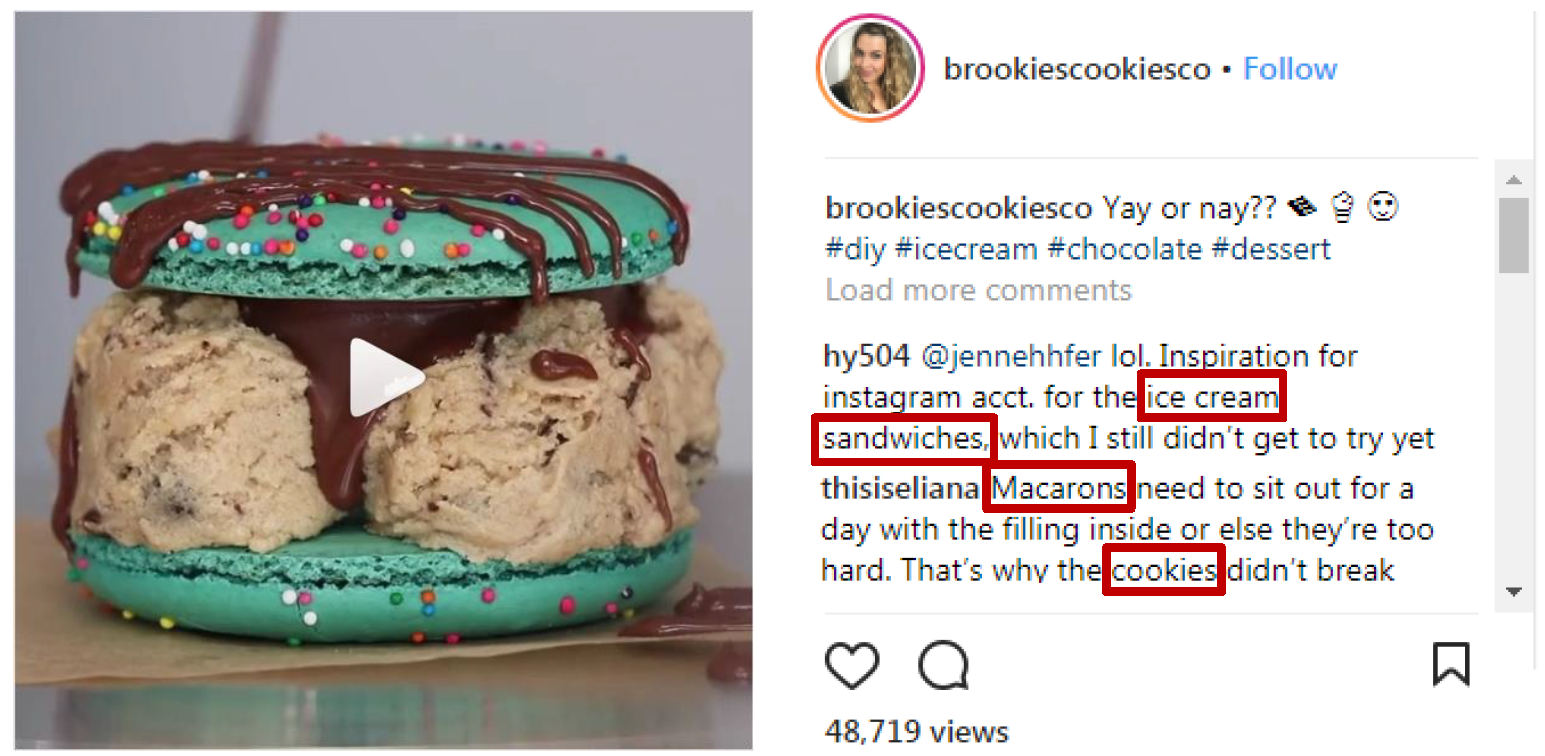}
	  \vspace{-1mm}
	  \caption{Demonstration of user interests reflected by the comments. The red boxes are the entity words.}
	\label{fig_1}	  
	\vspace{-5mm}
\end{figure}
\section{Related Work}
Our work is related to a broad spectrum of thumbnail selection and cross-modal similarity computation.
\subsection{Thumbnail Selection}
As mentioned above, the thumbnail selection approaches can be divided into two categories. The mono-modal thumbnail selection approaches focus on learning the visual
representativeness purely from the visual content. For instance, Kang et al.~\cite{3} defined the concept of the “representativeness” and divided the criterion into four main attributes: frame quality, visual details, content dominance, and attention measurement to identify the thumbnail. To choose the representative frame, Guan et al.~\cite{4} proposed a top-down approach to represent each frame with global features via utilizing a scalable clustering method. And then, they chose the frame that best covered a set of local descriptors with minimal redundancy. Further, Lu et al.~\cite{5} suggested that the importance of individual local features should be exploited to
calculate the representativeness. They presented a novel Bogof-Importance (BoI) model to compute the representativeness of each frame by aggregating the weighted importance of the local features contained in the frame. However, these methods often fail to produce satisfying results as they neglect the side-information associated with the videos. Some sideinformation, such as the title and description from publishers,
always can be used to represent the content of the video. Others from the users, such as comments and queries, bridge the users and videos.

Therefore, recently, some researchers have started to investigate and leverage the side-information for video understanding. Several multi-modal video thumbnail selection
approaches have been proposed. Gao et al.~\cite{6} incorporated a web image set, corresponding to keywords of videos, into the thumbnail selection method. Liu et al.~\cite{7} matched the query words and the frame content to raise a query-specific
thumbnail selection algorithm. To bridge the preference gap between the video owners and browsers, Zhang et al.~\cite{8} proposed a unified framework for web video thumbnail
recommendation by analyzing the relationship between the video thumbnails and browsers’ search queries. Liu et al.~\cite{9} learned a joint embedding space of the visual
and textual information to propose another query-specific thumbnail selection method. Nevertheless, these query-specific thumbnail selection methods still pay attention to the representativeness calculation of the query information and ignore the users’ interests. Moreover, these methods are not applicable to the current micro-video platforms for they do not have the searching operation. In contrast, our proposed approach focuses on locating the attractive thumbnail for each micro-video using the visual quality, representativeness, and popularity.
\section{DATA PREPARATION}
According to this, we detail the data preparation, including the auxiliary dataset, micro-video dataset construction, and popular topics collection.
\subsection{Auxiliary Dataset}
To train the deep transfer model, the auxiliary dataset should be large-scale, high-quality, and adaptive to our crossmodal multi-label similarity calculation. Overall, Open Images Dataset\footnote{http://storage.googleapis.com/openimages/web/index.html} is adopted as training dataset. It is a large-scale dataset with 9,011,219 images mainly collected from the Flickr\footnote{https://www.flickr.com/}, and each image is annotated with multiple labels over 600 categories. The labels cover the real-life entities and
are manually drawn by professional annotators to ensure the accuracy and consistency.
\subsection{Micro-video Dataset Construction}
To build the set of micro-videos, lack of the ground-truth of each micro-video still poses a big challenge. Previous studies choose the ground-truth thumbnails by some workers, but this strategy is inadaptive to our task. In our method, we suggest that the thumbnail is the frame whose content is largely related to the popular topics of the microvideo shared platform, yet the topics are affected by the realworld users’ interests which are probably different from those annotating workers’. Therefore, we collect the micro-videos with a large number of clicks to build the dataset and treat their posted thumbnails as the ground-truths. Because these posted thumbnails are the first sights of the micro-videos and convey the only information about the content. More clicks on the thumbnail mean it satisfies more users and is closer to the popular topics of the platform.

In practice, we collect some publishers and capture their posted micro-videos associated with the thumbnails. Since the aim is to obtain the micro-videos with a large number of
clicks, whose thumbnails are close to the popular topics, the influence of publishers themselves has to be eliminated. For this, we choose the publishers according to the number of their followers. As can be seen in Figure 2, the number of followers is limited to the range of 300-1000. Ultimately, we achieve 18,279 publishers to form the publishers set and crawl their micro-videos with more than 200 clicks.

To improve the performance of the thumbnail selection, some candidate frames are extracted from each micro-video based on visual quality and representativeness. Specifically, according to the findings, three types of low-quality frames would be detected and filtered out, including frames of dark, blurry, and uniform-colored. And we further filter out the near-duplicate frames via a clustering algorithm over the
high-quality frames. As shown in Figure 2, most micro-videos have 7-12 candidate frames. Although the posted thumbnail may not be contained in the set of candidate frames, we believe its nearest neighbor frame in the set is visually closer to the posted thumbnail. Statistically, the average similarity between the posted thumbnail and its nearest neighbor frame is about 93.8\%. Therefore, we take the nearest neighbor frames as the ground-truths for the micro-videos, instead of inserting the posted thumbnails into the sets of candidate frames. To verify the ground-truth, 1, 000 groups of candidate frames are randomly selected and annotated by 10 Instagram active users. From Figure 6, we observe that with the increasing number of clicks, the accuracy of manual labeling is improved. The number of clicks reflects the degree of attractiveness of the thumbnail to the users. Moreover, the accuracy can exceed 60\% in the worst case, proving that the ground-truth is more suitable than other frames as the thumbnail.
    \begin{figure}
	\centering
	  \includegraphics[width=0.4\textwidth]{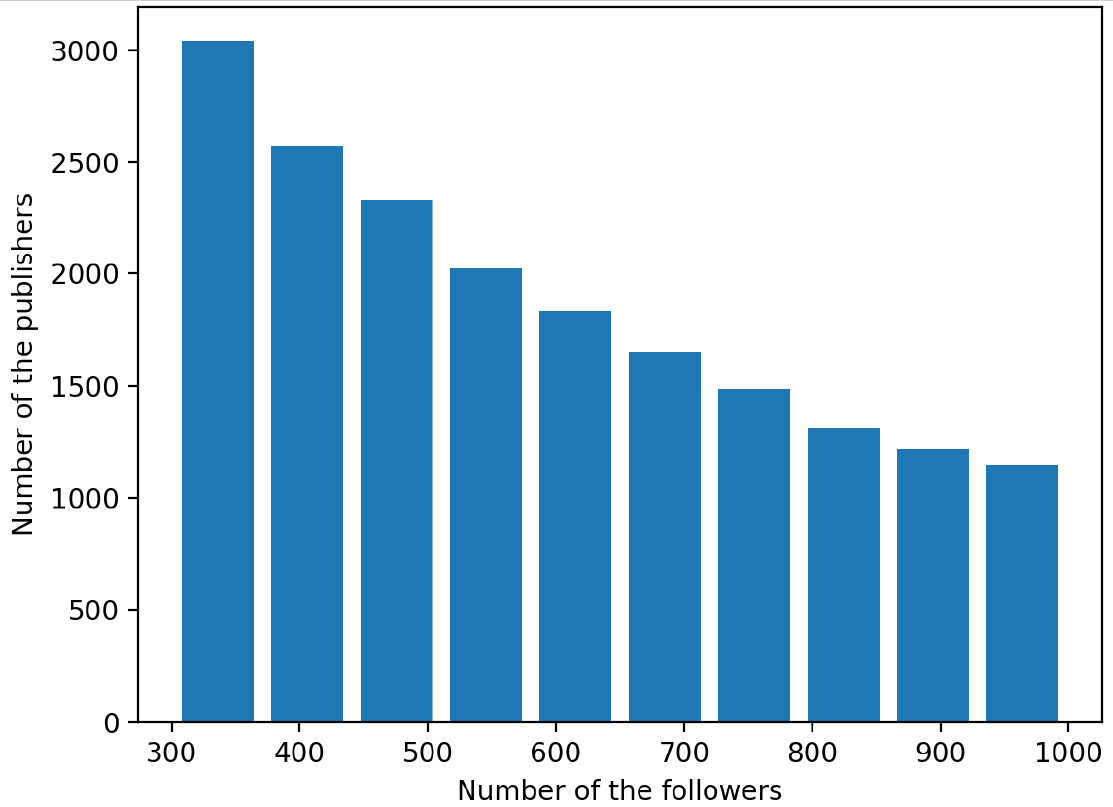}
	  \vspace{-1mm}
	  \caption{Distribution of the number of followers in our dataset.}
	\label{fig_4}	  
	\vspace{-5mm}
\end{figure}

\subsection{Popular Topics Collection}

After the micro-video dataset collection and ground-truth construction, we focus on the popular topics collection. Therefore, about 1 million applicable comments are crawled
from Instagram to capture popular topics. Finally, we go through the entity words and obtain 111 high-frequency words to represent the popular topics. Table I lists a part of entity words and their frequencies.

\begin{table}
  \centering
  \caption{List of the representative popular topics and their frequencies in our dataset.)}
  \label{table_1}
  \setlength{\tabcolsep}{2.0mm}
  \begin{tabular}{|c|c|c|c|}
    \hline
    \textbf{Word}&\textbf{Times}&\textbf{Word}&\textbf{Times}\\
    \hline
    Girl&29211&Beach&2388\\
    \hline
    Baby&15873&Cake&2225\\
    \hline
    Boy&1048&Guitar&2122\\
    \hline
    Car&8783&Football&2013\\
    \hline
    Dog&7230&Child&1982\\
    \hline
    Gym&7001&Chocolate&1790\\
    \hline
    Family&6154&Bike&1678\\
    \hline
    Smoke&4873&Bag&1524\\
    \hline
  \end{tabular}
  \vspace{-2mm}
\end{table}
\section{EXPERIMENTS}
In this part, we carried out extensive experiments to thoroughly validate our proposed model and its components on the constructed micro-video dataset.

\subsection{Parameter Settings}

We implemented our model with the help of PyTorch\footnote{https://pytorch.org/}. For the deep transfer model, we employed the AlexNet-based deep transfer model introduced in~\cite{36}. Notably, the model extends the AlexNet architecture comprised of five convolutional layers (conv1 - conv5) and three fully connected layers (fc6 - fc8). The pre-trained AlexNet is adapted, whose conv1-conv3 layers have been frozen and the conv4 - conv5 layers would be fine-tuned at the training stage. Besides, we applied Xavier approach to initializing the model parameters, proven as an excellent initialization method for the neural networks. 
Following the prior work~\cite{HUIGN}, the mini-batch size and learning rate are respectively searched in \{128, 256, 512\} and \{0.0001, 0.0005, 0.001, 0.005, 0.01\}. The optimizer is set as Adam. Moreover, we empirically set the size of each hidden layer as 256 and the activation function as ReLU. Without special mention, except the deep transfer network, other models employ one hidden layer and one prediction layer. For two embedding space construction, we explored the dimensionality of them in \{50-500\} and \{1024, 2048\}, respectively. For a fair comparison, we initialized other competitors with the analogous procedure. We showed the average result over five-round predictions in the testing set.
    \begin{figure}
	\centering
	  \includegraphics[width=0.4\textwidth]{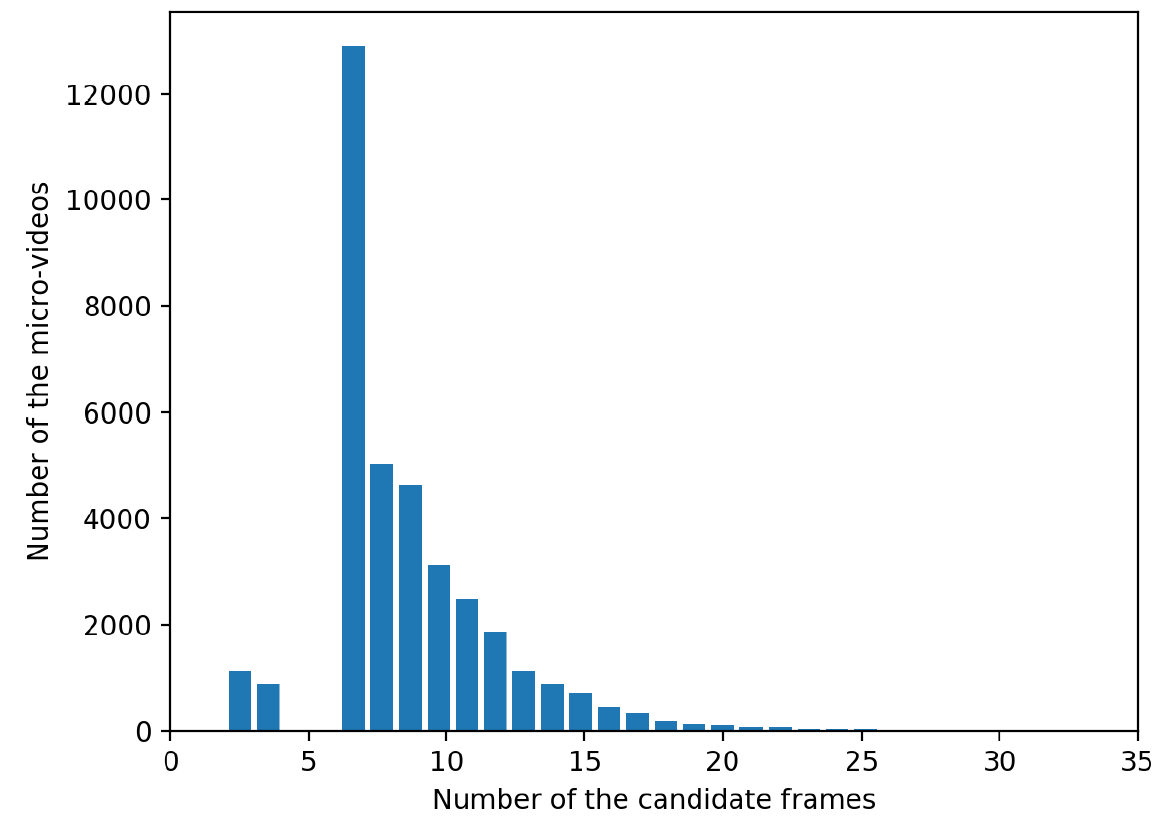}
	  \vspace{-1mm}
	  \caption{The distribution of the extracted candidate frames in our micro-video dataset.}
	\label{fig_5}	  
	\vspace{-5mm}
\end{figure}

\subsection{Baselines}

To evaluate our proposed model for the thumbnail selection, we compared it to the following six methods on the labeled micro-video dataset:
\begin{itemize}
    \item Random. The method randomly selects one frame from the candidate frame set as the thumbnail.
    \item ASBTV~\cite{26}. The authors proposed an automatic thumbnail selection system to exploit the representation and the visual attractiveness of frames. This system measures the
attractiveness by analyzing the visual quality and aesthetic metrics, as well as the representativeness achieved by their proposed clustering algorithm.
    \item DeViSE~\cite{25}. The model is the first to construct a semantic embedding space through the natural language model and map the visual information into this space to resolve the scalability problem of standard recognition models. We leveraged this model to select an attractive thumbnail for each micro-video via popularity computation.
    \item MTL-VSEM~\cite{9}. The authors developed a multi-task deep visual-semantic embedding model to automatically select the query-dependent video thumbnails according to both the visual and side-information. We retrained this model to compute the relevance between the frame and popular topics, instead of the visual information and queries.
    \item C2AE~\cite{35}. The authors proposed a novel deep neural network (DNN) based end-to-end model for solving multilabel classification task. This model integrates the DNN architectures of the canonical correlation analysis and autoencoder, allowing the learning and prediction with the ability to exploit label dependency.
\end{itemize}
\subsection{Performance Comparison}
The comparative results are shown in Table I. From this table, we have the following observations:
\begin{itemize}
    \item  In terms of accuracy, the random method performs the worst, indicating the significance of the video content.
    \item  The side-information based models (e.g., MTL-VSEM and C2AE) outperform ASBTV which only considers the visual quality and representativeness. This result demonstrates that the side-information can improve the micro-video understanding.
    \item  When performing the thumbnail selection task, MTLVSEM is superior to DeViSE. It is reasonable since MTL-VSEM employs the multi-task learning which refers to the joint training of multiple tasks, while enforces a common intermediate parameterization or representation to improve each task’s performance.
    \item  C2AE surpasses the visual-semantic embedding based approach DeViSE. This verifies that the multi-label task benefits our thumbnail selection, since the popularity calculation should consider the similarities between the pairs of each frame and all words in the list of popular topics. And the standard visual-semantic embedding model causes the more significant error on the distances calculation to multiple prototypes.
    \item  It is observed that MTL-VSEM outperforms C2AE. The discrepancy of the training data and testing data is the primary cause, while C2AE ignores the gap between them, and MTL-VSEM integrates the multi-task learning to decrease this discrepancy. In addition, C2AE cannot be extended to the unseen prototypes, leading to the external error.
\end{itemize}
\begin{table}
  \centering
  \caption{Performance comparison between our model and
the baselines.}
  \label{table_2}
  \setlength{\tabcolsep}{2.0mm}
  \begin{tabular}{|c|c|}
    \hline
    \textbf{Method}&\textbf{Accuracy}\\
    \hline
    \textbf{Random}&$14.09\pm 2.81\%$\\
    \hline
    \textbf{ASBTV}\cite{26}&$20.92\pm 0.31\%$\\
    \hline
    \textbf{DeViSE}\cite{25}&$24.09\pm 0.35\%$\\
    \hline
    \textbf{MTL-VSEM}\cite{9}&$27.01\pm 0.57\%$\\
    \hline
    \textbf{C2AE}\cite{35}&$26.16\pm 0.53\%$\\
    \hline
  \end{tabular}
  \vspace{-2mm}
\end{table}

     \begin{figure*}
	\centering
	  \includegraphics[width=1\textwidth]{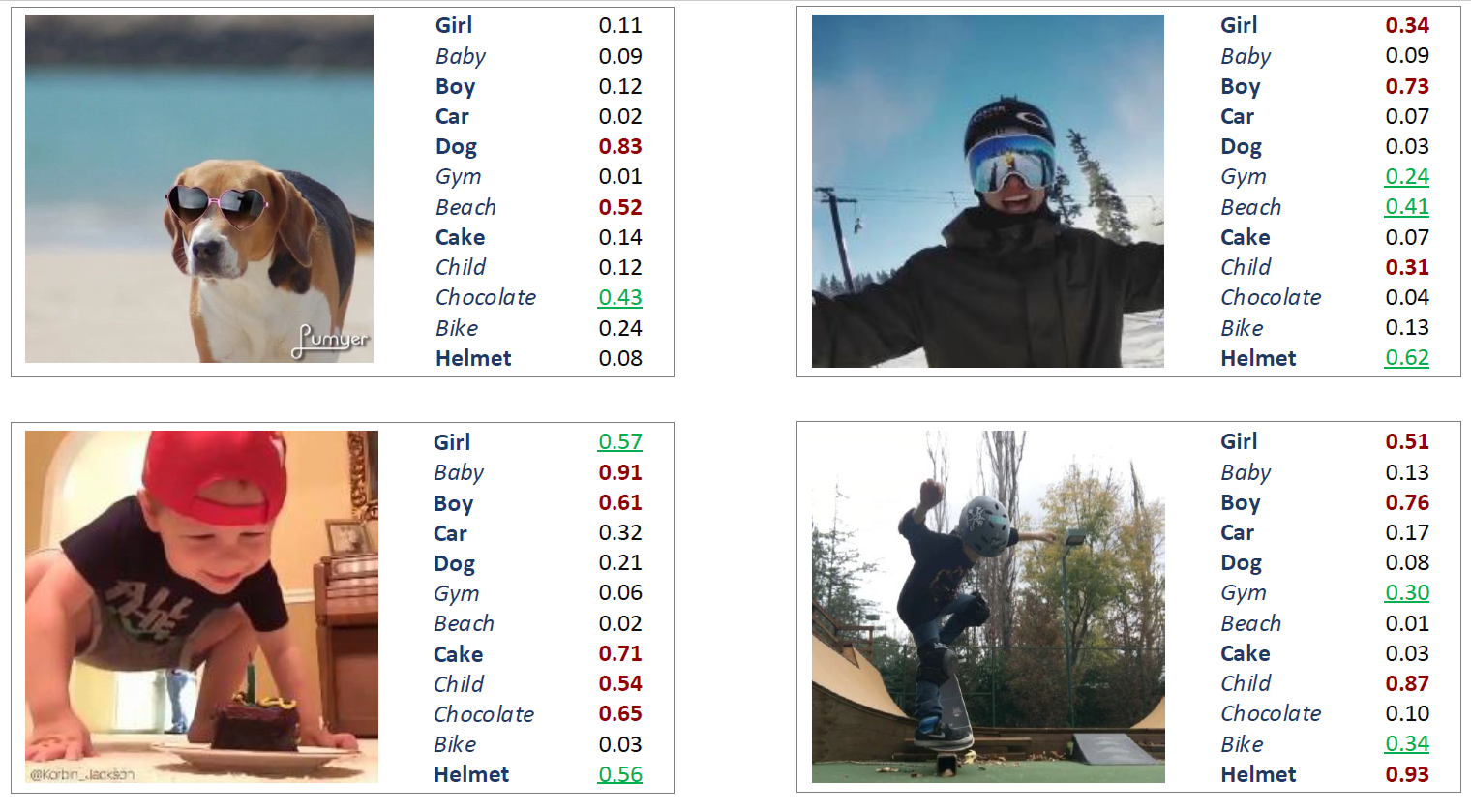}
	  \vspace{-1mm}
	  \caption{Visualization of the similarity scores between the frame and popular topics.}
	\label{fig_10}	  
	\vspace{-5mm}
\end{figure*}

\subsection{Visualization}

Following the exiting work~\cite{PHRec}, we showed several examples drawn from our model to visualize the similarity between the images and popular topics. As shown in Figure 3, there are four visualization experiment groups. Each group consists of a frame extracted from a micro-video and some representative words, where the bold ones are the seen labels in the Open Images Dataset and the italic ones are unseen, from popular topics associated with the similarity scores computed by our proposed model. The bold red scores denote the relatively higher similarity and the underlined scores in green means that the frame has little relevance to the words with overvalued scores. For instance, in the first group, the words “Dog” and “Beach” have higher ratings than the others. It is in line with the correlation between the frame and the words. Also, the score of the unseen word “Chocolate” is far from the true similarity. However, the similarity of this word in other groups can be calculated correctly. Further analysis, the scores of the seen words, such as “Dog”, “Boy” and “Cake”, are more likely to coincide with the true correlations. Despite the unseen words “Chocolate”, “Bike” and “Gym” tend to cause errors, the unseen words “Child” and “Baby” are accurate. Besides, the seen words “Helmet” and “Girl” seems to fail in giving the exact scores sometimes, yet we observed that the mistakes happened to the highly visual similar objects. Above all, we suggested that the proposed model can be used to calculate the similarity between the frame and the seen/unseen words.

\section{Conclusion and Future Work}
In the future, we expect to construct a new method for thumbnail selection and  investigate a personalized thumbnail recommendation model which can yield the different thumbnails for the users.
\bibliography{BIB/IEEEabrv, reference}

\end{document}